\documentstyle[12pt]{article}
\begin{document}


\vspace{0.5cm}
\begin{center}

{\Large
STRONG AND RADIATIVE MESON DECAYS
\smallskip
IN A GENERALIZED
\smallskip
NAMBU--JONA-LASINIO MODEL
  \footnote{Work supported in part by Deutsche Forschungsgemeinschaft,
Schweizerischer Nationalfonds,  JNICT no. PMCT/C/CEN/72/90, GTAE and CERN no.
PCERN-FAE-74-91.}}

\vspace{1.5cm}

V. Bernard$^{\dag}$, A.H. Blin$^{\ddag}$, B. Hiller$^{\ddag}$,
U.-G. Mei\ss ner$^{\S}$\footnote{Heisenberg fellow} and M.C. Ruivo$^{\ddag}$

\vspace{0.5cm}

$^{\dag}$Physique Th\'eorique, Centre de Recherches Nucl\'eaires et
Universit\'e Louis Pasteur de Strasbourg. B.P. 20, F-67037 Strasbourg
Cedex 2, France

$^{\ddag}$Centro de F\'{\i}sica Te\'orica (INIC), Departamento de F\'{\i}sica
da Universidade, P-3000 Coimbra, Portugal

$^{\S}$Institut f\"ur Theoretische Physik, Universit\"at Bern, Sidlerstrasse 5,
CH-3012 Bern, Switzerland

\end{center}
\vspace{3cm}

{\sl
We investigate strong and radiative meson decays in a generalized
Nambu--Jona-Lasinio model. The one loop order calculation provides a
satisfactory
agreement with the data for the mesonic spectrum and for radiative decays.
Higher order effects for strong decays  of $\rho$ and $K^*$
are estimated to be large.
We also discuss the role of the flavour mixing determinantal
interaction.
}
\vfill

\pagebreak

\section{Introduction}
\vspace{1cm}

The Nambu--Jona-Lasinio model [1] and generalizations thereof
      [2,3,4] have
been used extensively to study the properties of mesons in free space and
at finite temperatures and densities. It is an effective field theory of
non-linearly interacting quarks which exhibits spontaneous and explicit
dynamical chiral symmetry breaking. In the case of three flavours, it is
mandatory to incorporate the 't Hooft six-fermion interaction to describe
the breaking of the axial $U(1)$ symmetry       [3]. Mesons are bound quark-
antiquark pairs in this approach and their properties can readily be
calculated by solving the pertinent Bethe-Salpeter equations. However, no
systematic study of three-point functions like the strong and radiative
meson decays, has been performed so far. Previous
attempts were limited to
the leading term in the momentum expansion of the underlying quark-meson
vertices [5] and thus do not account for the full dynamics of the model.
Furthermore, these decays are also a good testing ground to find out the
limitations of the model. This will be discussed in some detail later on.

\vspace{1cm}

In what follows, we will work in flavour $SU(3)$ and use the following
Lagrangian

\begin{eqnarray}
{\cal L}=& &G_{1}[(\bar{\psi} \lambda_i \psi)^{2}
+(\bar{\psi} i \gamma_{5} \lambda_{i} \psi)^{2}] \nonumber \\
 &+&G_2[(\bar{\psi} \lambda_{a} \gamma_{\mu} \psi)^{2}
+(\bar{\psi} \lambda_{a} \gamma_{\mu} \gamma_{5} \psi)^{2}] \nonumber \\
 &+&K[{\rm det}\{\bar{\psi} (1+\gamma_{5}) \psi\}+{\rm det}\{\bar{\psi}
(1-\gamma_{5}) \psi\}]
\label{e1}
\end{eqnarray}
\noindent
where the flavour index $i$ runs from 0 to 8 with
$\lambda_{0}=\sqrt{2/3} \, {\bf 1}$, the $\lambda_a$ are color
matrices $(a=1,...,8)$. As it stands, the
Lagrangian is characterized by a few
parameters: The two four--fermion coupling constants $G_1$ and $G_2$, the
six--fermion
coupling $K$ and the cutoff $\Lambda$, which is necessary to regularize the
divergences. We will use a covariant four-momentum cutoff $\Lambda = 1$ GeV.
Furthermore, to account for the explicit symmetry breaking, a quark mass
term has to be added. We work in the isospin limit $m_u = m_d$  and will use
the current quark masses to fit the meson spectrum. Clearly, the coupling
$G_1$
is related to the properties of the pseudoscalar Goldstone bosons, $G_2$ can
be fixed from the $\rho$-meson mass and $K$ is necessary to give the $\eta
\eta'$ mass
splitting. This completely specifies the model and we are now at a point to
consider its dynamical content.

\section{Formalism}

The basic object to consider is the triangle diagram which describes the
coupling of the decaying meson $(M_1)$ into the other mesons $(M_{2,3})$
 or another
meson $(M_2)$  and a photon  $\gamma$ or two photons $\gamma_1$, $\gamma_2$
via the quark loop. Let us
first consider the strong decays. Dropping all prefactors, the transition
amplitude for the process $M_{1} \rightarrow M_{2} M_{3}$ can be evaluated by
working out (cf. fig.1)
\begin{equation}
\Gamma(M_{1} \rightarrow M_{2} M_{3}) = Tr (\Gamma_{M_{1}} S_{F} \Gamma_{M_{2}}
S_{F} \Gamma_{M_{3}} S_{F})
\label{e2}
\end{equation}
where $\Gamma_{M_{i}}$ gives the $i^{\rm th}$
meson-quark-antiquark vertex and
$S_F$ the propagator
of the constituent  quarks. The latter follows from minimizing the
effective potential to one loop. The Bethe-Salpeter vertex functions
relevant for our considerations are of scalar, pseudoscalar and vector type

\begin{displaymath}
\Gamma_{S} = g_{S} {\bf 1} \otimes I
\end{displaymath}
\begin{equation}
\Gamma_{P} = g_{P} (1+h_{P} \not{p}) \gamma_{5} \otimes I
\label{e3}
\end{equation}
\begin{displaymath}
\Gamma_{V} = g_{V} \gamma_{\mu} \otimes I
\end{displaymath}
\noindent
Here, $I$ is a generic symbol for the isospin structure and we have
introduced scalar, pseudoscalar and vector $M {q \bar{q}}$-couplings. The
coupling $h_{p}$
stems from the pseudoscalar-axial vector meson mixing. This is discussed in
more detal in refs. [2,4,6,7]. Since we work in the isospin limit, no
scalar-vector mixing arises in the $SU(2)$ subgroup. In the scalar and
pseudoscalar channels, a further complication is due to the $\lambda_{0}
\lambda_{8}$ mixing
which has already been discussed in ref.[8] in some detail. The solution
of the corresponding Bethe-Salpeter equations is standard and we do not
exhibit any details here.

Let us briefly elaborate on the connection between the various transition
amplitudes and meson-meson coupling constants. Consider first the decay of
a scalar $(S)$ into two pseudoscalars $(P)$. The transition amplitude is a
purely scalar function, called $T_{S}$, and we have
\begin{displaymath}
\Gamma(S \rightarrow P P) = \frac{|\vec{p}_{c}| |T_{S}|^{2}}{8 \pi E_{S}^{2}}
\end{displaymath}
\begin{equation}
G^{2}_{SPP} = \frac{|T_{S}|^{2}}{4 \pi}
\label{e4}
\end{equation}
with $|\vec{p}_{c}|$ the momentum of an outgoing particle in the rest frame
of the
decaying particle, $|\vec{p}_{c}| = ((s-(m_{1}+m_{2})^{2})(s-(m_{1}-m_{2})
^{2})/4s)^{1/2}$
and $s=m^{2}(M_{1})$. For the decay of a vector $(V)$ into two
pseudoscalars, one has

\begin{displaymath}
T_{\mu}(V \rightarrow PP) = (p_{1}-p_{2})_{\mu} G_{VPP}
\end{displaymath}
\begin{equation}
\Gamma (V \rightarrow PP) = \frac{|\vec{p}_{c}|^{3} G_{VPP}^{2}}{6 \pi
m_{V}^{2}}
\label{e5}
\end{equation}

\noindent
Finally, for the reaction $V \rightarrow \tilde{V} P$ we find

\begin{displaymath}
T_{\mu \nu}(V \rightarrow \tilde{V} P) = \epsilon_{\alpha \beta \mu \nu}
p^{\alpha}_{\tilde{V}} q^{\beta} {\cal F}
\end{displaymath}
\begin{equation}
\Gamma(V \rightarrow \tilde{V} P) = \frac{|\vec{p}_{c}|^{3}}{3} \,
\frac{{\cal F}^{2}}{4 \pi}
\label{e6}
\end{equation}

In the case of the radiative decays, one can use the same formalism since
the photon behaves much like a vector particle. Since our Lagrangian
contains no tensor interaction term, the pertinent photon-quark-antiquark
vertex takes the minimal form

\begin{equation}
\Gamma_{\gamma} = \frac{e}{2} (\lambda_{3} + \frac{1}{\sqrt{3}} \lambda_{8})
\label{e7}
\end{equation}

For the decay of a pseudoscalar into two photons, we have a structure
similar to the one for $V \rightarrow \tilde{V} P$, the only difference
being that in the formula
for the width $\Gamma (P \rightarrow \gamma \gamma)$ one has a factor $1/2$
instead of $1/3$ which is the
reduction in plolarization degrees of freedom for a massless particle.

\section{Estimates of higher order effects}

The description of mesons as $q\bar{q}$ pairs has been proven
to be quite successful
in the calculation of the meson mass spectrum.
 However it may not account for other properties of some mesonic resonances,
such as their decays.
 We anticipate that this is indeed the case for the strong decays
$\rho \rightarrow \pi\pi$
and $K^* \rightarrow K\pi$, which come out very small in the one loop
calculation.
 The insufficiency of a $q\bar{q}$ description of the mesons was already
emphasized by
Krewald et al [9] in the context of the pion electromagnetic form
factor.

Due to the failure of the one loop approximation for the
calculation of those decays, it is necessary
to estimate the magnitude of
higher order effects, such as the two loop corrections. The full calculation
is,
for the moment, out of the scope of the present work.
 A simple estimate of such effects for the mesonic decays can be obtained
by calculating a dressed meson propagator, as shown in fig.2.
The dashed loop refers to $\pi\pi$ or $K\pi$ states, in the case of the $\rho$
and
$K^*$ propagators, respectively. By bare propagator we denote the meson
described as a $q\bar{q}$ state.

The full propagator reads

\begin{equation}
G_{\alpha\,\beta}\,=G^{0}_{\alpha\,\beta}\,+\,G^{0}_{\alpha\,\lambda}\,\,
\Sigma_{\lambda\,\mu}\,\,G_{\mu\,\beta}
\label{e8}
\end{equation}

\noindent where the bare propagator is

\begin{equation}
G^{0}_{\alpha\,\beta}\,=\,\frac{g_{\alpha\,\beta}\,-\,{\hat q}_\alpha\,\,
{\hat q}_\beta}
{q^{2}\,\,-\,m^{2}_V}\,\,=\,\,\frac{T_{\alpha\,\beta}}
{q^{2}\,\,-\,m^{2}_V}
\label{e9}
\end{equation}

\noindent and  $\Sigma_{\lambda\,\mu}$ is the meson loop given by

\begin{equation}
\Sigma_{\lambda\,\mu}\,\,=\,\,8\,G^{2}_{VPP}\,\,\, \int_{\Lambda_1}
\frac{d^{4}k
}{(2\pi)^{4}}\,\,
((2k-q)_{\lambda})\,((2k-q)_{\mu})\,S(k,m_{P1})\,S(k-q,m_{P2})
\label{e10}
\end{equation}

\noindent with $S(k,m_{Pi})$ the propagators of the mesons
obtained in the one loop order, $m_{P1}$ and $m_{P2}$
being their masses. The remaining factors in the integrand correspond to the
coupling of the vector mesons to the pseudoscalars in
the first order calculation (5). The covariant cutoff $\Lambda_1$ needs not to
be the same as $\Lambda$. One obtains:

\begin{equation}
G_{\alpha\,\beta}\,=\,\frac{T_{\alpha\,\beta}}{q^{2}\,\,-\,m^{2}_V
\,-\,C}
\label{e11}
\end{equation}

\noindent with $C\,\equiv C\,(\,\,G^{2}_{VPP}\,\,,q^{2}\,\,,m_{P1}\,\,,
m_{P2}\,\,,\Lambda_1\,\,)$.
The quantity C has a cut for $q^2\,>(m_{P1}\,+\,m_{P2}\,)^2$ and one rewrites
$G_{\alpha\,\beta}$ as

\begin{equation}
G_{\alpha\,\beta}\,=\,T_{\alpha\,\beta}\,\,\frac{Z}{q^{2}\,\,-\,{\tilde
m_V}^{2}\,-\,i\,{\rm Im}\,C\,({\tilde m_V}^{2}\,)\,Z}
\label{e12}
\end{equation}

\noindent provided that the renormalization factor $Z$ is roughly constant
around the physical mass
$\tilde m_V$ and where the renormalization factor is:

\begin{equation}
Z\,=\,(\,1\,-\,\frac{d}{d\,q^2}\,\,{\rm Re}\,C\,)^{-1}\,
\biggl|_{q^2\,=\,{\tilde m_V}^2}
\label{e13}
\end{equation}

The decay width of the vector meson is finally

\begin{equation}
\Gamma\,\,(V \rightarrow PP)\,=\,\frac{{\rm Im}
\,C\,({\tilde m_V}^2)\,Z}
{{\tilde m_V}}.
\label{e14}
\end{equation}

Using this scheme the amplitudes for radiative decays of $\rho$ and $K^*$ have
then to be also multiplied
by $\sqrt{Z}$ (wave function renormalization).

\section{Results and discussion}

At the one loop level we use the meson spectrum and decay constants to fix the
parameters of the
model. For $\Lambda =1$ GeV, $G_{1} \Lambda ^{2} = 3.95, G_{2} \Lambda ^{2}
= 5.43,
K \Lambda ^{5} = 42, m_{u} = m_{d} = 4$ MeV and $m_{s} = 115$ MeV,
we find the following
meson masses (the experimental values are given in parentheses for
comparison): $M_{\pi} = 136.5 (139.6)$, $M_{K} = 497.5 (497.7)$,
$M_{\eta} = 549 \, (548.8)$, $M_{\eta'} = 936 \, (957.5)$,
$M_{\rho} = 775 \, (768.3)$, $M_{\omega} = 764 \, (781.95)$,
$M_{K^*}  = 898 \, (891.6)$, $M_{\Phi} = 990 \, (1019.41)$ and
$M_{a_{0}} = 970 \, (983.3)$ (all in MeV). For the
pion and kaon decay constants, we have $F_{\pi} = 93.9$ MeV and $F_K = 96.6$
MeV, i.e. the ratio $F_{K}/F_{\pi}$
is too small, which is a common feature in this kind of models. We find an
overall satisfactory description of the meson spectrum together with
reasonable values for the vacuum expectation values of the scalar quark
densities $\bar{u} u$ and $\bar{s} s$, $-<\bar{u}u>^{1/3}=272$ MeV $
(225\pm 35)$ and $<\bar{s}s>/<\bar{u}u> = 0.74 (0.8\pm 0.2)$.

In table 1, we show the results for the strong decays in
comparison to the
empirical values.
Obviously, for states in
the quark-antiquark continuum the results are not reliable as indicated by
the decay $\Phi \rightarrow \pi \rho$.
Also, for our set of parameters the decay $\Phi \rightarrow
\bar{K}K$ is kinematically
forbidden. The large width of the $\Phi \rightarrow \pi \rho$ decay is due
to the too strong
flavour mixing induced by the six-fermion interaction proportional to $K$.
This could presumably be cured by including more terms in the Lagrangian
like e.g. in ref.[6].

As for the strong decays of $\rho$  and $K^*$ the one loop order calculation
is clearly insufficient to account for the respective empirical widths.
Using the simple
 approximation
 scheme, described in the previous section, to include the second order
effects
and approximating the meson propagators in the loop by propagators of
structureless particles, the results improve by about  a factor of 2.
We think, therefore, that it is mandatory to consider a more complex
multiquark structure  for the $\rho$ and $K^*$ mesons. The strong $\rho$
decay width
including second order effects is still quite small as compared to the
experimental one,
but this number should be understood only as a guide for the order of magnitude
of higher loop corrections.  We notice that the parameters
could have been adjusted in order to have the correct decay
width for the $\rho$ and the KSFR relation fulfilled, but  at the cost of
having bad values for the radiative decays.

Let us now turn to the radiative decays. In table 2, our results are
summarized. We find an overall satisfactory description of the data,
the main exceptions being the widths for $\Phi \rightarrow \eta \gamma$ and
for the $K^* \rightarrow K \pi$. In the first case this
is, again, an artifact of the interaction Lagrangian used and also, since the
$\Phi$ lies in the
unphysical quark-antiquark continuum, should not be considered
significantly troublesome.
The fact that the ratios $\Gamma_{VPP}^{calc} / \Gamma_{VPP}^{exp}$
and $\Gamma_{VP\gamma}^{calc} / \Gamma_{VP\gamma}^{exp}$
 are larger for the $K^*$ decays than for the $\rho$ decays (both strong and
radiative)
 is consistent with the small ratio $F_K / F_\pi$.
       We have also investigated the case $K=0$ (no six-fermion term).
Reducing the strange quark mass to $80$ MeV [2], which is necessary
to find a
decent fit to the spectrum, one is not able to get a satisfactory
description of strong and radiative decay widths.
After finishing this work, we became aware of a preprint by Takizawa
and
Krewald [10], who deal with the radiative decays $\pi^0 \to \gamma
\gamma$
and $\eta  \to \gamma \gamma$ in a similar model. Their Lagrangian contains
the four--fermion interaction proportional to $G_1$ and the six--fermion
determinantal interaction. While their results are similar to ours, we
disagree with their conclusions at various places. First, in the case of
$\pi^0 \to \gamma \gamma$ they remove the cut--off to find agreement with the
current algebra prediction. This is, however, not a consistent procedure
since once the cut--off is introduced, the effective theory is defined  and
should not be altered in the process of calculating various quantities.
Second, for calculating the width of $\eta  \to \gamma \gamma$, they
use the
empirical $\eta$ mass, which is 17 per cent larger than the value they find
within the model. This, of course, alters substantially the result for this
particular width. Comparing their results with  ours, we also find a
satisfactory description for these two particular radiative decays. It should
be obvious, however, that the model is somewhat too crude too draw as far
reaching conclusions as done in ref.[10]. As long as one is not able to
properly account for the $SU(3)$ breaking effects in the pseudoscalar decay
constants, it is doubtful that one can make firm quantitative statements
about such breaking effects in other processes.

In summary, we have used the generalized three-flavour NJL model to calculate
strong and radiative meson decays (three-point functions) taking
the full solution to the Bethe-Salpeter equations in the one-loop
approximation. A reasonable agreement to the experimental data is obtained,
with the exception of the $\rho$ and $K^*$   decays. This seems to indicate
that
 a more complex
multiquark structure should be accounted for these mesons. This conjecture is
supported
by a simple estimate of the two loop order corrections.
We also demonstrated the importance of the flavour-mixing determinantal
six--fermion interaction. Further studies including also the effects of
isospin--breaking seem necessary to understand some fine details of
the mesonic
interactions at low energies and to gain insight into effects of $SU(3)$
breaking on various observables.

\bigskip

\bigskip

TABLES
\vspace{1cm}

\begin{table}[h]
\begin{center}
\begin{tabular}{|l|ccccc|} \hline
 &$\rho \rightarrow \pi \pi$&$K^{*+} \rightarrow \pi^{+} K^{0}$&$
K^{*+} \rightarrow \pi^{0} K^{+} $&$ a_{0} \rightarrow \pi \eta $&$
\phi \rightarrow \pi \rho$ \\ \hline
NJL(I)&52.0&18.0&9.0&74.5&1.5 \\ \hline
NJL(II)&94.0&38.2&19.1&-&- \\ \hline
Exp.&$151.5 \pm 1.2$&$38.6 \pm 0.6$&$16.7 \pm 0.3$&$57 \pm 11$&$0.6 \pm 0.3$
\\ \hline
\end{tabular}
\caption[]{Strong meson widths for various decays in units of MeV:
 (I) calculated in  one loop order; (II) with estimates of two loop order
included.}
\label{t1}
\end{center}
\end{table}
\vspace{1cm}

\begin{table}[h]
\begin{center}
\begin{tabular}{|l|cccc|} \hline
 &$\pi ^{0}\rightarrow \gamma \gamma$&$\eta\rightarrow \gamma \gamma$&$
\rho ^{\pm}\rightarrow \pi ^{\pm}\gamma$&
$ \rho \rightarrow \eta \gamma $\\ \hline
NJL &$7.9\cdot 10^{-3}$&0.77&60.1&60.4  \\ \hline
Exp.&$(7.7\pm 0.6) \cdot 10^{-3}$&$0.46\pm 0.04$&$67.1\pm 7.6$&
$57.6\pm 10.7$ \\ \hline
 &$\omega \rightarrow \pi^{0}\gamma $&$\omega \rightarrow \eta \gamma $&$
K^{*+}\rightarrow K^{+} \gamma$&$\Phi\rightarrow\eta\gamma$  \\ \hline
NJL&762&6.3&92.0&259.0  \\ \hline
Exp.&$716.6 \pm 43.0$&$4.0\pm 1.9$&$50.3\pm 4.6$&$56.7\pm 2.8$  \\ \hline
\end{tabular}
\caption[]{Anomalous and radiative meson decay widths in units of keV,
calculated in the one loop order.
 The second order corrected decays are $\Gamma_{\rho\pi\gamma}\,\,=\,63$ keV
and $\Gamma_{K^*K\gamma}\,\,=\,98.9$ keV ($\Lambda_1=1.3$ GeV).}
\label{t2}
\end{center}
\end{table}

\bigskip

FIGURE CAPTIONS

\bigskip

Fig. 1: Quark triangle diagram to calculate the strong meson decays. In the
case of radiative decays, one has to substitute the third meson $M_{3}$ by
a photon, and for the anomalous decays  $M_{2}$ and $M_{3}$ by two photons.

\bigskip

Fig. 2: The dressed vector meson propagator (thick line) includes a
2-pseudoscalar excitation (dashed line), which is a 2nd order effect. The one
loop order ($q\bar q$ state) is denoted by the thin line, the bare meson
propagator.
\end{document}